\documentclass[sigconf]{acmart}

\usepackage[normalem]{ulem}

\usepackage{balance}

\newcommand{\hatm}[1]{\expandafter\hat#1}

\usepackage{mathtools}
\usepackage{multirow}

\definecolor{myred}{rgb}{0.7,0.1,0.1}
\definecolor{mygreen}{rgb}{0.1,0.5,0.1}

\newcommand{\ignore}[1]{}

\usepackage{pifont}
\newcommand{\cmark}{{\color{mygreen}\ding{52}}}
\newcommand{\xmark}{{\color{myred}\ding{55}}}

\usepackage{xspace}
\usepackage{tabularx}

\newcommand{\shrinkAroundFigure} {\vspace{-6mm}}
\newcommand{\shrinkAroundTable}  {\vspace{-3mm}}
\newcommand{\shrinkAroundCaption}  {\vspace{-4mm}}
\newcommand{\shrinkAroundSection}  {\vspace{-4mm}}

\begin{document}

\title{GENIEx: A \underline{G}eneralized Approach to \underline{E}mulating \underline{N}on-\underline{I}d\underline{e}ality in Memristive \underline{X}bars using Neural Networks}
\author{Indranil Chakraborty, Mustafa Fayez Ali, Dong Eun Kim, Aayush Ankit and Kaushik Roy}
\email{ichakra@purdue.edu}
\affiliation{%
  \institution{School of Electrical and Computer Engineering, Purdue University}
  \streetaddress{465 Northwestern Avenue}
  \city{West Lafayette}
  \state{Indiana}
  \postcode{47907}
}

\begin{abstract}

 Memristive crossbars have been extensively explored for deep learning accelerators due to their high on-chip storage density and efficient Matrix Vector Multiplication (MVM) compared to digital CMOS.
 However, their analog nature of computing poses significant issues due to various non-idealities such as: parasitic resistances, non-linear I-V characteristics of the memristor device etc.
 The non-idealities can have a detrimental impact on the functionality i.e. computational accuracy of crossbars.
 Past works have explored modeling the non-idealities using analytical techniques.
 However, several non-idealities have data dependent behavior.
 This can not be captured using analytical (non data-dependent) models thereby, limiting their suitability in predicting application accuracy.

 To address this, we propose a Generalized Approach to Emulating Non-Ideality in Memristive Crossbars using Neural Networks (GENIEx), which accurately captures the data-dependent nature of non-idealities.
 First, we perform extensive HSPICE simulations of crossbars with different voltage and conductance combinations.
 Based on the obtained data, we train a neural network to learn the transfer characteristics of the non-ideal crossbar. 
 Next, we build a functional simulator which includes key architectural facets such as \textit{tiling}, and \textit{bit-slicing} to analyze the impact of non-idealities on the classification accuracy of large-scale neural networks.
 We show that GENIEx achieves \textit{low} root mean square errors (RMSE) of $0.25$ and $0.7$ for low and high voltages, respectively, compared to HSPICE.
 Additionally, the GENIEx errors are $7\times$ and $12.8\times$ better than an analytical model which can only capture the linear non-idealities.
 Further, using the functional simulator and GENIEx, we demonstrate that an analytical model can overestimate the degradation in classification accuracy by $\ge10\%$ on CIFAR-100 and $3.7\%$ on ImageNet datasets compared to GENIEx.
 
\end{abstract}

\date{}
\maketitle

\thispagestyle{empty}

\shrinkAroundSection

\section{Introduction}
% 1. Emergence of specialized accelerators for deep learning, and those using emerging memory technologies

The pervasiveness of deep learning in a wide-variety of applications such as object detection, language processing etc. has been a major force behind the recent success of Artificial Intelligence (AI).
Consequently, there has been a growing interest in developing specialized accelerators to improve the efficiency of deep learning.
Such accelerators include Google TPU~\cite{jouppi2017datacenter}, Microsoft BrainWave~\cite{chung2018serving}, and Nvidia V100.
One key aspect driving these accelerators is moving computations closer to the memory, which has brought forth the paradigm of in-memory computing.
Despite the breakthroughs in custom hardware, the storage and computation requirements of Deep Neural Networks (DNNs) have been increasing at a much faster rate than the efficiency improvements in digital CMOS hardware~\cite{xu2018scaling}. 
To this effect, researchers have explored Non Volatile Memory (NVM)~\cite{Hu2018MNIST, ambrogio2018equivalent} based crossbar architectures to achieve higher on-chip storage density and efficient MVMs in the analog domain~\cite{shafiee2016isaac, ankit2019puma}.

% 2. Introduce NVMs, their non-ideality and categorize them (linear->non data-dependent/analytical, non-linear->data-dependent)

NVM devices can store multiple states per device, and crossbars built with these devices can be integrated on chip leading to high storage density~\cite{Hu2018MNIST}.
Second, the voltage-driven nature of these two-terminal devices enables crossbar-like arrangement to perform MVMs, at significantly higher efficiency compared to digital CMOS~\cite{hu2016dac}.
Despite the multifold promises of NVM technologies, the analog nature of computing in crossbars poses several challenges due to the device and circuit non-idealities such as: parasitic resistance, non-linearity from access transistors, and I-V characteristics of the NVM device.
Parasitic resistances lead to undesirable IR-drops in the metal lines of the crossbar.
On the other hand, the non-linearity leads to inaccurate multiplications at the cross-points.
As a result, non-idealities can have an adverse effect on the MVM arithmetic.
This gets exacerbated further due to the device variations.
Eventually, the inaccuracies in the MVM arithmetic can accumulate over the multiple layers of a neural network, causing significant accuracy degradation~\cite{jain2019cxdnn}.

% 3. Introduce past work, and GENIEx as a means to overcome past issues

To address this accuracy degradation, there have been efforts towards exploring techniques to model non-idealities and subsequently mitigating them  \cite{jain2019cxdnn,  chakraborty2018technology, agrawal2019x}. 
The efficacy of these  mitigation techniques strongly depend upon the  modelling \cite{jain2019cxdnn, chakraborty2018technology, jeong2017parasitic} approach to exhaustively capture the sources of the non-idealities and retraining of the neural network weights.
The non-idealities in crossbars can be broadly categorized into non-data dependent or linear (for eg. parasitic resistances), and data-dependent or non-linear types (for eg. access transistors and device I-V characteristics). 
While the current analytical techniques can model the non-data dependent aspects \cite{jain2019cxdnn,  chakraborty2018technology, jeong2017parasitic}, they fail to capture the data dependent non-idealities.
Data-dependent non-idealities can have a pronounced effect on the crossbar outputs, particularly at higher operating voltages (discussed in Section \ref{sec:analysis}).
Thus, it is important to move away from approximate analytical models to data-based models in order to truly capture all the non-idealities. 
In this work, we present GENIEx, a neural network based modelling approach that provides an accurate as well as generalized representation of the non-ideal behavior of crossbars.
The key contributions of this work are:

\vspace{-1mm}
\begin{itemize}
    \item Analyze the sources of non-ideality in crossbars through extensive SPICE simulations (Section~\ref{sec:analysis}).
    \item Propose GENIEx, a generalized approach for modelling non-idealities in crossbars using neural networks (Section~\ref{sec:nn-model}).
    \item Develop a PyTorch-based functional simulator which models the key architectural aspects namely tiling, and bit-slicing to evaluate large-scale DNNs using GENIEx (Section~\ref{sec:func-sim}).
    \item Perform detailed analysis of different non-idealities on the classification accuracy of DNNs (Section~\ref{sec:results}).
\end{itemize}
\vspace{-1mm}

To the best of our knowledge, this is the first work proposing an end-to-end framework for data-dependent crossbar modeling along with a functional simulator considering tiling, and bit-slicing.
This enables studying the accuracy impacts of device and circuit properties at the application level.
It is worth noting that due to the ability to capture data dependency of crossbar behavior (transfer characteristics), GENIEx can be used to model crossbars from both simulations as well as experimental measurements. 
\textit{We believe that our proposed approach paves the way for universal modeling of practical crossbars with the scope of seamless functional evaluation and mitigation.}
We plan to open-source the framework for further research on crossbar hardware.

\section{Related Work}
\label{sec:related}
Past research have explored modeling crossbar non-idealities and subsequently mitigating them~\cite{liu2014reduction, jain2019cxdnn, chakraborty2018technology, agrawal2019x}. 
Jain et al~\cite{jain2019cxdnn} used matrix inversion techniques to model the effects of parasitic resistances due to input driver, metal lines \textit{etc}. 
Liu et al~\cite{liu2014reduction} proposed an approximation technique based on sample input/output behavior. 
An alternative way of capturing effects such as stuck-at-faults~\cite{liu2017rescuing} or device variations~\cite{liu2015vortex} is to map the distribution of the variations or defects. 
While the above modelling approaches~\cite{jain2019cxdnn, liu2017rescuing, liu2014reduction, liu2015vortex} consider linear (non-data dependent) non-idealities, GENIEx also captures the non-linear (data dependent) non-idealities. 
Note, there could be non-linearity during programming of NVM devices. 
Sun et al~\cite{sun2019impact} propose analytical models to study such non-linearity during programming. 
However, analyzing the impact of non-linearity on the subsequent MVM computations (after programming) requires a data-dependent model like GENIEx. 

Researchers have also proposed evaluation frameworks such as~\cite{jain2019cxdnn} and NeuroSim~\cite{chen2018neurosim} to study the impact of these non-idealities using analytical models. 
Other works have explored the impact of quantization noise of ADCs for analog computing~\cite{rekhi2019analog}.
However, these frameworks do not consider the architectural aspects of MVM computations such as tiling and bit-slicing, which have a significant implication on classification accuracy (shown in Section~\ref{sec:results-quantization}). 
Our work explores a neural network based technique to model the crossbar non-idealities using a functional simulator with detailed MVM architecture. 
Table~\ref{tab:related-table} summarizes our contribution with respect to the related work.

\begin{table}[t]
    %\shrinkAroundTable
    \caption{Related work comparison}
    \shrinkAroundCaption
    \centering
        \resizebox{1.0\columnwidth}{!}{
            
\begin{tabular}{|l|c|c|c|}
\hline
\textbf{Related Work}                                           & \multicolumn{1}{l|}{\textbf{\begin{tabular}[c]{@{}l@{}}Linear + Non-linear \\ non-idealities\end{tabular}}} & \multicolumn{1}{l|}{\textbf{\begin{tabular}[c]{@{}l@{}}Large scale \\ DNNs\end{tabular}}} & \multicolumn{1}{l|}{\textbf{\begin{tabular}[c]{@{}l@{}}Architecture \\ model of MVM\end{tabular}}} \\ \hline
\hline
\textbf{GENIEx}                                                 & \cmark                                                                                       & \cmark                                                                     & \cmark                                                                              \\ \hline
\textbf{CxDNN~\cite{jain2019cxdnn}}       & \xmark                                                                                       & \cmark                                                                     & \xmark                                                                              \\ \hline
\textbf{CrossSim~\cite{CrossSim}}         & \cmark                                                                                       & \xmark                                                                     & \xmark                                                                              \\ \hline
\textbf{NeuroSim~\cite{chen2018neurosim}} & \cmark                                                                                       & \xmark                                                                     & \xmark                                                                              \\ \hline
\textbf{AMS~\cite{rekhi2019analog}} & \xmark                                                                                       & \cmark                                                                     & \xmark                                                                              \\ \hline
\end{tabular}
        }
    %\shrinkAroundTable
    \label{tab:related-table}
    \vspace{-6mm}
\end{table}

%A popular way of mitigating these non-ideal effects in large-scale neural networks is to retrain the weights using the analytical models in the forward pass \cite{jain2019cxdnn}. In addition, there have been efforts on modifying the backpropagation algorithm \cite{chakraborty2018technology} to account for these errors. Other techniques involve weight remapping algorithms \cite{agrawal2019x}.

%Most of the aforementioned works focus on linear or distribution based modeling which capture the effect of the static resistances or device variations. However, memristive device characteristics can be non-linear where the non-linearity is often voltage dependent \cite{guan2012spice}.  However, inference operations can also be affected by non-linear I-V characteristics. Moreover, access transistors can add further non-linear effects in the computations. %Thus, there is a need to look at models which can capture the data dependent nature of the non-idealities and generalize across the input space of large-scale neural networks.

\section{Analysis of NVM Non-Idealities} \label{sec:analysis}

\begin{figure}[t]
	\centering
	\includegraphics[width=0.85\columnwidth, keepaspectratio]{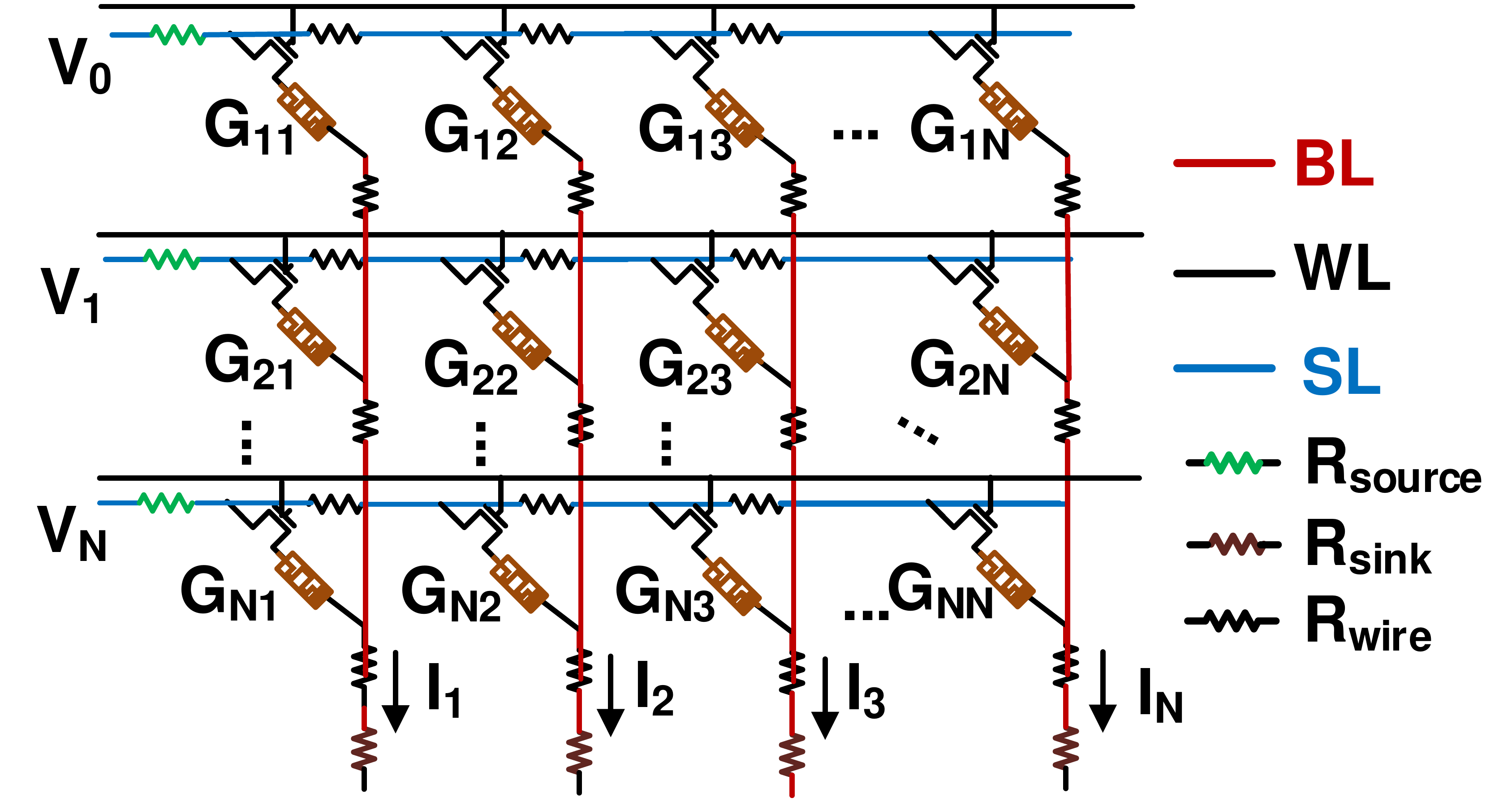}

	\caption{A typical non-ideal crosspoint structure with NVM devices accompanied by a transistor at every junction of the word-lines (WL) and bit-lines (BL).}
	\label{fig:crossbar}
	\vspace{-4mm}
\end{figure}

\textbf{Background: }
A typical memristive crossbar consists of NVM devices arranged in a crossbar fashion as shown in Figure \ref{fig:crossbar}. 
The two terminals of each NVM device connect to a horizontal word-line (WL) and a vertical bit-line (BL). 
These devices are accompanied by access transistors or selectors to avoid the sneak path issues during writing \cite{zidan2013memristor}. 
This primitive can be used to compute Matrix Vector Multiplications (MVMs) in the \textit{analog domain} by activating all the WLs and sensing all the BLs simultaneously. 
For example, to perform a multiplication between a $1\times N$ vector and a $N\times M$ matrix, the vector is encoded as input voltages~($V_i$) while the matrix is encoded as conductances~($G_{ij}$).
Consequently, the output current in the $j^{th}$ BL (\textit{for ideal crossbar}) is the sum of currents through each NVM device in the corresponding column: $I_{j} = \sum_i V_iG_{ij}$.
Thus, the currents from the $M$ columns constitute the output vector of the MVM operation. 
Typically, a crossbar requires peripheral circuits such as Digital-to-Analog Converters (DACs) and Analog-to-Digital Converters (ADCs) for system-level integration.
The DACs convert the digital inputs into analog voltages while the ADCs convert the analog currents in the BLs to digital outputs.
Due to the analog nature of computing, several non-idealities can lead to errors in the MVM computations. 
These non-idealities can be classified into two kinds - linear and non-linear, as shown in Table~\ref{tab:nonid-table}.

\begin{table}[h]
    \shrinkAroundTable
    \caption{Non-idealities in crossbar}
    \vspace{-3mm}
    \centering
        \resizebox{0.85\columnwidth}{!}{
            
\begin{tabular}{|c|c|}
    \hline
    \textbf{Linear Non-idealities}            & \textbf{Non-linear Non-idealities}   \\ \hline
    \hline
    Source Resistance ($R_{source}$) & Access devices or selectors \\ \hline
    Sink Resistance ($R_{sink}$)   & Device non-linearity        \\ \hline
    Wire Resistance ($R_{wire}$)     &                             \\ \hline
\end{tabular}
        }
    \vspace{-3mm}
    \label{tab:nonid-table}
\end{table}

\textbf{Analysis: }
Under the influence of non-idealities, the crossbar design parameters such as size, ON resistance, conductance ON/OFF ratio etc. can have a considerable effect on the magnitude of errors in computations. 
To analyze this effect, we perform SPICE analysis of a $64\times64$ crossbar.
Herein, the linear non-idealities are modeled using parasitic resistances as shown in Figure~\ref{fig:crossbar}.
The access devices are based on transistor models from TSMC 65nm technology. 
The device models are adopted from a compact model of a filamentary RRAM~\cite{guan2012spice}, where the current flowing through the device can be expressed as: $I(d,V) = I_0 exp(\frac{d}{d_0})sinh(\frac{V}{V_0})$. 
Here, $d$ is the gap-size between the tip of the filament and electrode, $I_0$, $d_0$ and $V_0$ are fitting parameters. 

\begin{figure}[t]
	\centering
	\includegraphics[width=\columnwidth, keepaspectratio]{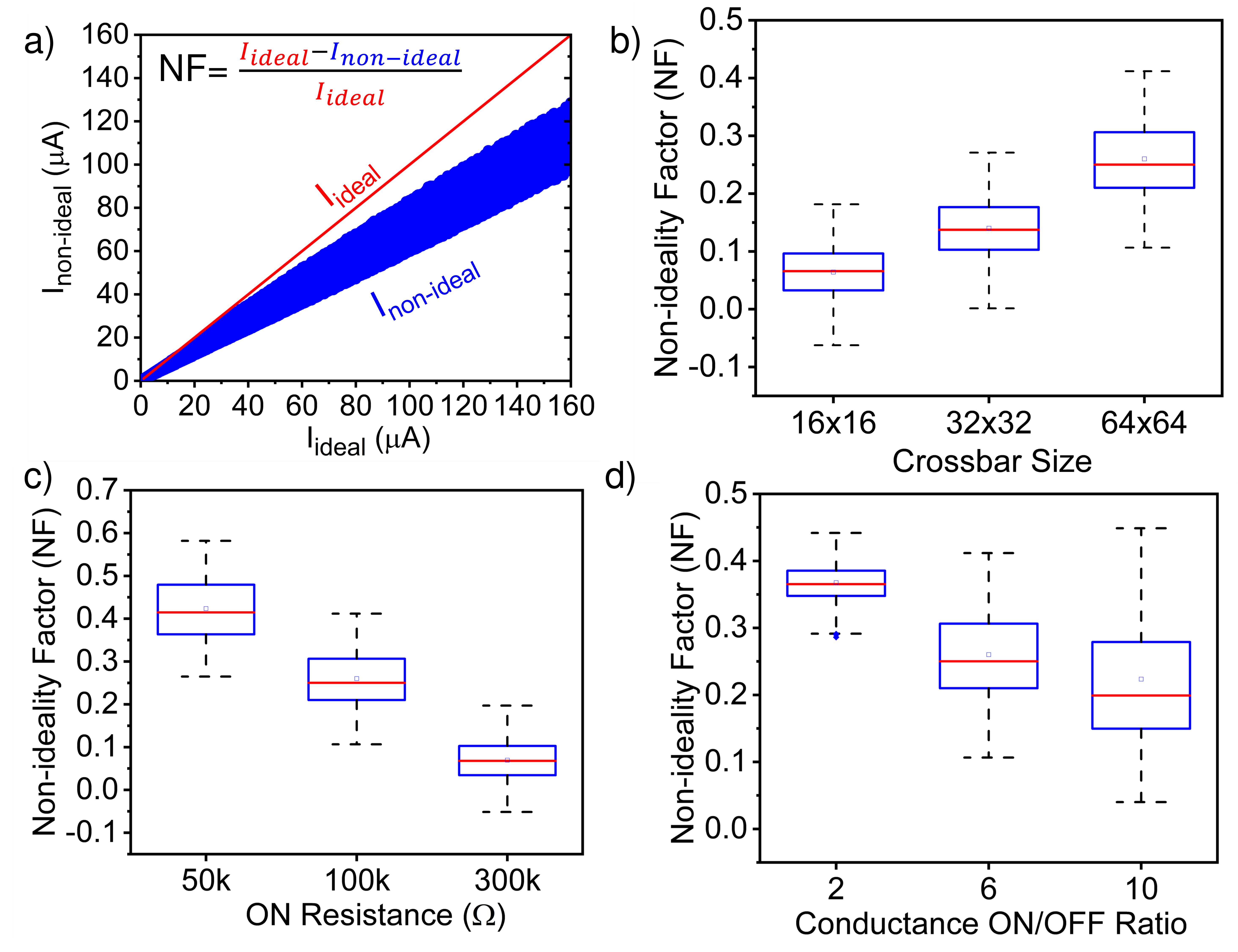}
    \vspace{-7mm}
	\caption{(a) Output currents from a 64x64 crossbar showing the deviation of ($I_{non-ideal}$) from ($I_{ideal}$). (b), (c) and (d) shows the box-plot variation of the NF with varying crossbar design parameters.  }
	\label{fig:nonideality}
	\vspace{-4mm}
\end{figure}

Figure \ref{fig:nonideality} (a) shows a typical plot of ideal current ($I_{ideal}$) v/s non-ideal current ($I_{non-ideal}$) of a crossbar.
Here, we observe that different voltage ($V$) and conductance ($G$) conditions which lead to similar $I_{ideal}$ can result in a varying range of $I_{non-ideal}$ outputs, causing errors in computations. 
To quantify the error, we define a non-ideality factor (NF) as the relative error between the $I_{ideal}$ and $I_{non-ideal}$. 
NF is calculated as: $\frac{I_{ideal}-I_{non-ideal}}{I_{ideal}}$. 
We observe in Figures~\ref{fig:nonideality} (b) and (c) that lower ON resistances and higher crossbar sizes lead to higher NF.
This is due to the fact that bigger crossbars have longer metal lines leading to higher $R_{wire}$. Moreover, the parallel combination of resistances along the columns and rows results in a reduced effective resistance of the crossbar in case of bigger crossbars as well as low ON resistances.
In addition, Figure~\ref{fig:nonideality} (d) shows that lower ON/OFF conductance ratio leads to high NFs. 
This is due to the fact that for a given ON resistance, the average resistance in the crossbar is low for lower $ON/OFF$ ratio. 
%It is thus necessary to model these crossbar level non-idealities to truly capture their effects on matrix vector multiplication computations and subsequently large scale workloads. 

\begin{figure}[t]
	\centering
	\includegraphics[width=\columnwidth, keepaspectratio]{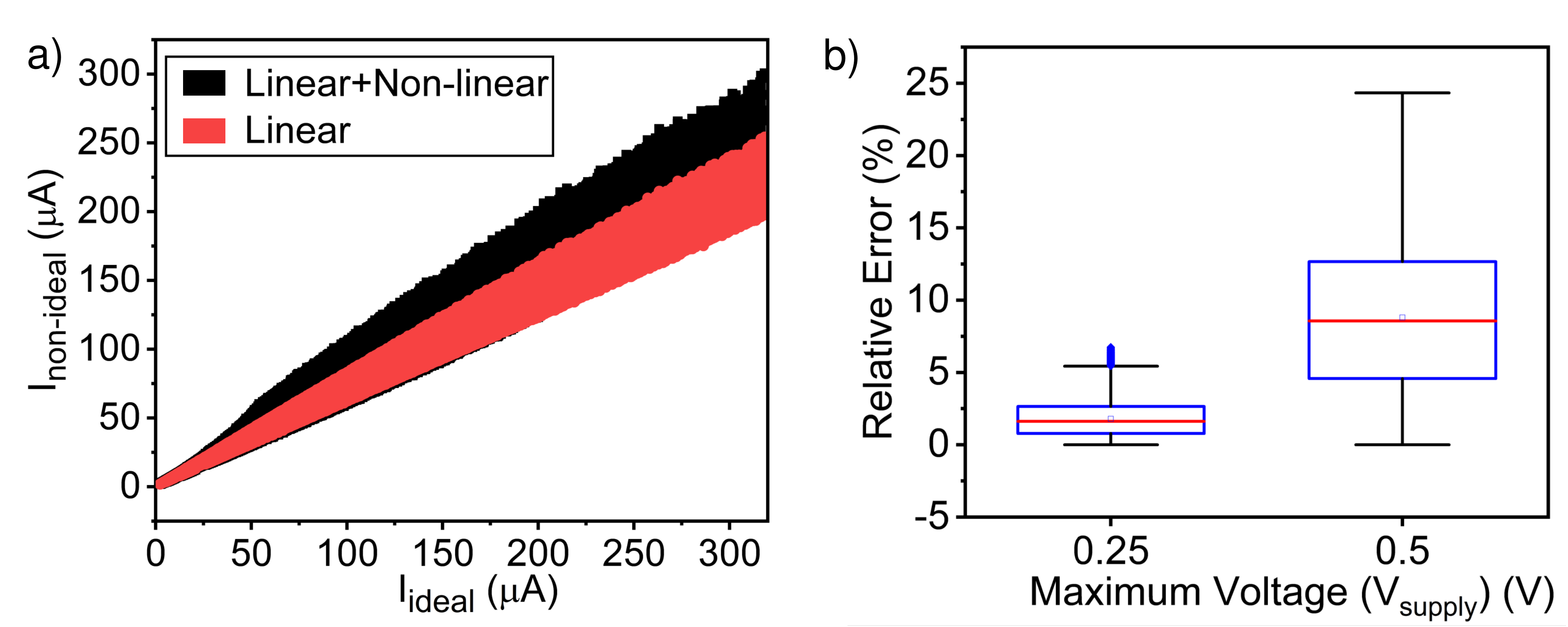}
    \vspace{-8mm}
	\caption{(a) Output current distribution showing impact of non-linearity. (b) Relative error between the cases with and without nonlinearity increases with increase in maximum supply voltage.}
	\label{fig:nonlinearity}
	\vspace{-4mm}
\end{figure}

Next, we analyze the impact of non-linear non-idealities. We consider two cases i) only linear non-idealities, ii) both linear and non-linear non-idealities.
Figures~\ref{fig:nonlinearity} (a) and (b) show the relative difference in output currents between the two cases.
We observe that the output currents in case (i) vary noticeably from case (ii). 
This effect becomes even more prominent for higher supply voltage of $V_{supply} = 0.5V$, thereby implying an inherent data dependence of $I_{non-ideal}$ on the $V$ and $G$.
\textit{This result underlines the drawbacks of analytical models which fail to capture the data-dependent non-idealities.}  
We propose a neural network based modeling technique that captures the data-dependent errors in crossbar computations.

%\clearpage

\section{GENIE\lowercase{x} - A Neural Network Based Crossbar Model} \label{sec:nn-model}

Neural networks project data to a high dimensional space which enables them to distinguish between different input patterns. 
We leverage this property of neural networks to propose GENIEx, which models the non-ideal behavior of memristive crossbars for different input voltage and conductance combinations. 
As discussed in Section~\ref{sec:analysis}, non-idealities in crossbars can lead to a varying range of $NF$ for similar $I_{ideal}$. 
Using a neural network can help us capture the data-dependent nature of such non-ideal behavior.

\textbf{NN Formulation:} The output current vector of an ideal crossbar ($I_{ideal}$) represents an MVM operation between $V$ and $G$.
Meanwhile, the output current vector from a real crossbar is non-ideal and can be expressed as a distorted MVM function: $I_{non-ideal} = f_D(V,G)$.
Therefore, it represents multiplicative behavior between the input variables, $V$ and $G$. 
The objective here is to model such non-ideality function $f_D(V,G)$ being input-dependent and having multiplicative behavior. 
The intuitive way of modeling $f_D(V,G)$ using neural networks is to provide $V$ and $G$ as inputs and obtain $I_{non-ideal}$ as output.
However, as neural networks perform linear transformations, it is difficult for them to model multiplicative interactions between its inputs. 
To avoid such input multiplications, we propose extracting only the distortion information of the real output current from $f_D(V,G)$.
We define a function which represents the ratio of $I_{ideal}$ to  $I_{non-ideal}$: $f_R(V,G) = \frac{I_{ideal}}{I_{non-ideal}}$. 
$f_R(V,G)$ represents the deviation of the $I_{non-ideal}$ from $I_{ideal}$, thus eliminating the need to capture multiplicative relationships. 
For an $N\times N$ crossbar, the input vector to the neural network is a concatenation of ($N\times 1$) voltage vector and ($N^2\times 1$) flattened conductance vector. 
The output vector obtained is $f_R(V,G)$ which is of size $N\times 1$. 
Subsequently, the $I_{non-ideal}$ is obtained using $I_{ideal}/f_R(V,G)$.

\textbf{Dataset:} To train GENIEx for predicting the ratio $f_R(V,G)$ for a set of $V$ and $G$ vectors, we create a dataset covering the exhaustive space of $V$ and $G$ combinations.
Crossbar-based accelerators commonly use bit-slicing to perform high precision MVM operations~\cite{shafiee2016isaac, ankit2019puma}.
We observed that this leads to high sparsity in $V$ and $G$ vectors across the popular deep learning tasks.
To exhaustively capture the resulting sparse data distributions, we consider various degrees of sparsity while generating the training set of $V$ and $G$.
We apply the $V$ and $G$ vectors to various crossbars and perform SPICE simulations to obtain the corresponding $I_{non-ideal}$. 
The obtained  $I_{non-ideal}$ is used to calculate $f_R(V,G)$, the prediction labels for the dataset. 
To evaluate the accuracy of GENIEx, we create a separate validation set of $V$, $G$ and expected $f_R(V,G)$.

\textbf{NN Topology:} GENIEx considers a two layer fully-connected neural network consisting of an input layer, a hidden layer and an output layer.
For a $N\times N$ crossbar, the size of the neural network is given as: $(N^2+N)\times P\times N$, where $P$ is the number of neurons in the hidden layer.
The training set mentioned above is used to train the neural network by feeding $V,G$ combinations as inputs and $f_R(V,G)$ as the output.

\begin{figure}[t]
	\centering
	\vspace{-4mm}
	\includegraphics[width=\linewidth, keepaspectratio]{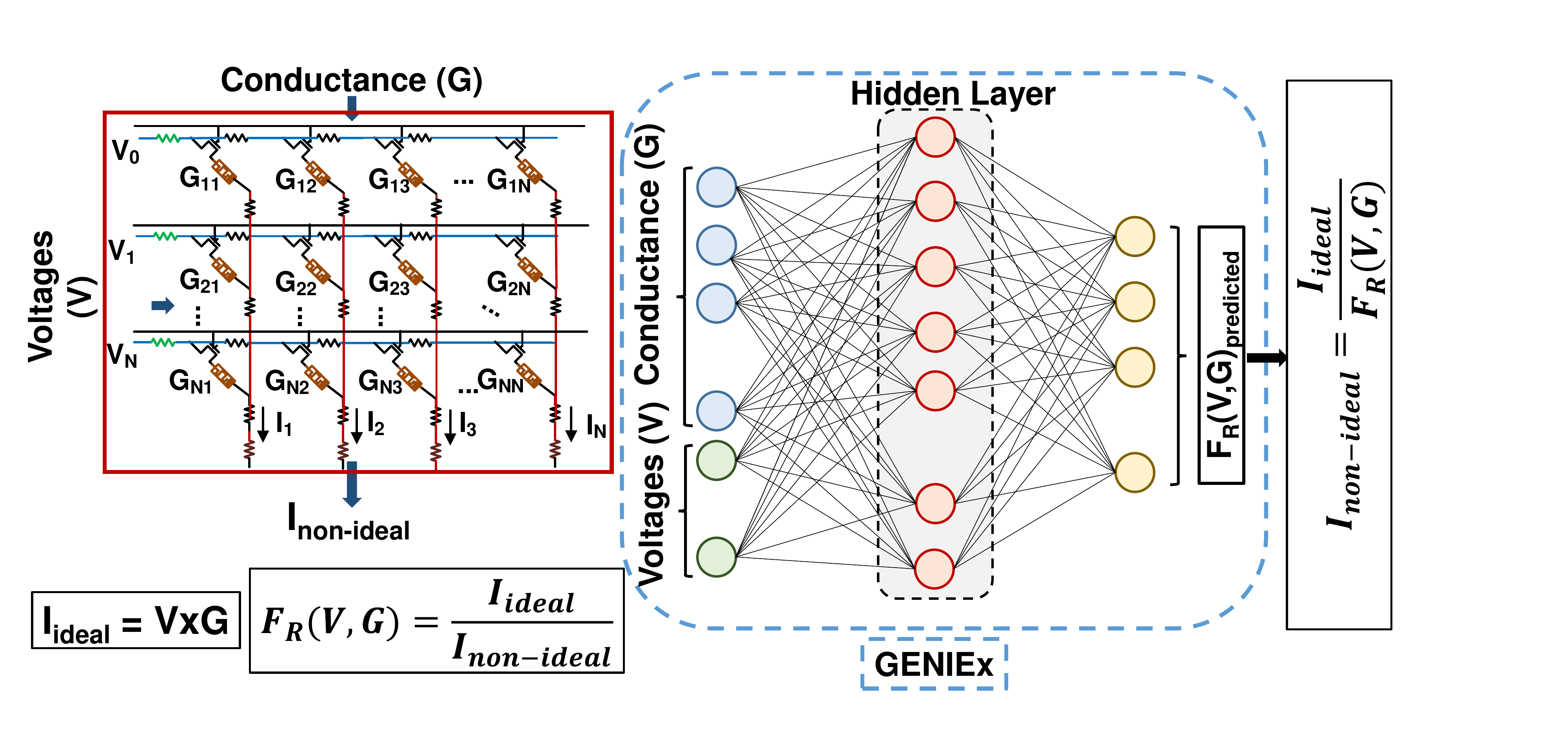}
	\vspace{-6mm}
	\caption{Crossbar computation mapped to GENIEx. $V$ and $G$ are concatenated to form the input vector for neural network, with output being the ratio $f_R = I_{ideal}/I_{non-ideal}$.}
	\label{fig:nn-model}
	\vspace{-6mm}
\end{figure}

\begin{figure}[t]
	\centering
	\includegraphics[width=\columnwidth, keepaspectratio]{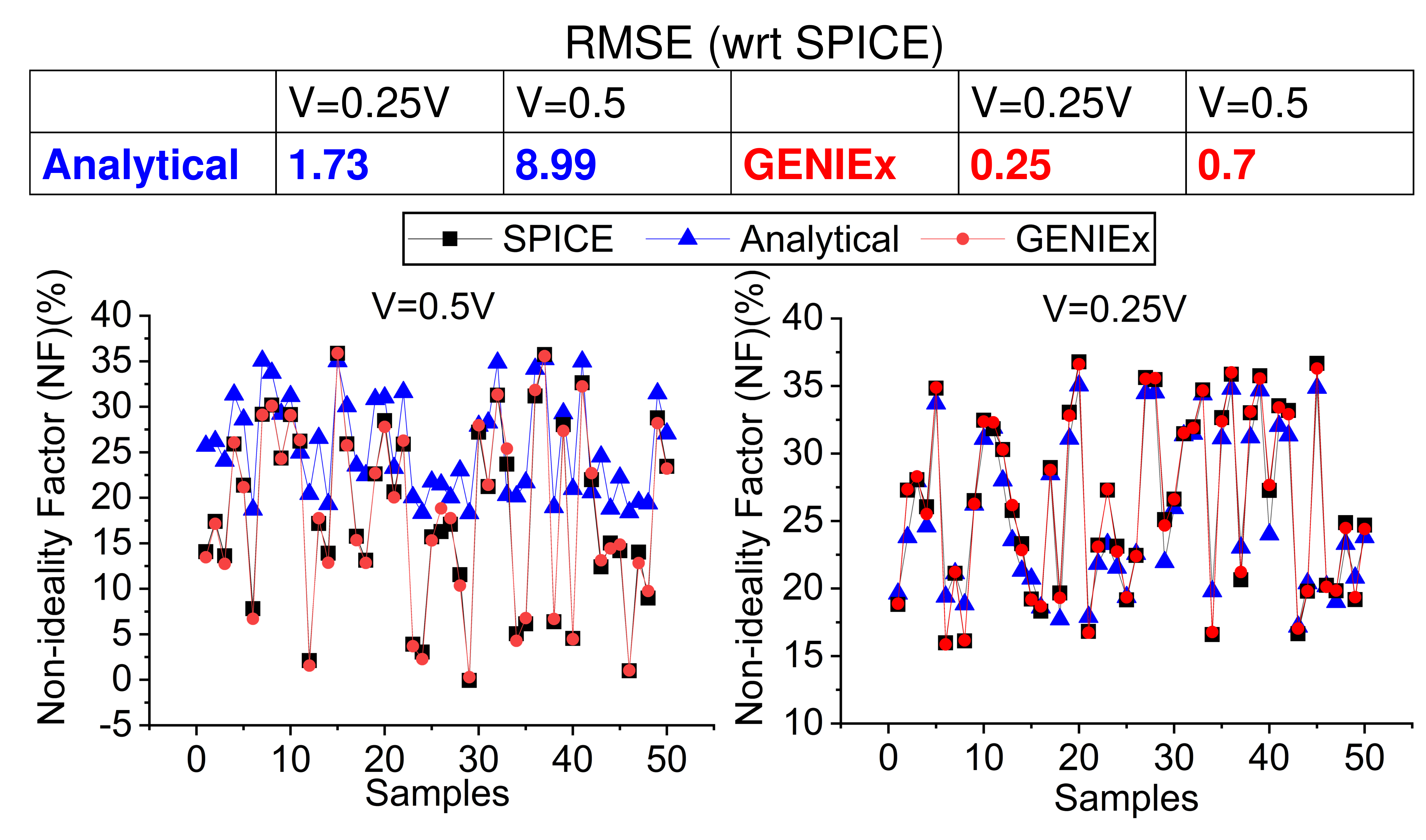}
    \vspace{-8mm}
	\caption{Comparison of $NF$ for a typical 64x64 crossbar between HSPICE outputs, analytical model and GENIEx.}
	\label{fig:model_v_ana}
	\vspace{-6mm}
\end{figure}

\textbf{Benchmarking: } We compare the accuracy of GENIEx against HSPICE results and a baseline linear analytical model for the same test voltage and conductance combinations. We use the metric $NF$, defined in Section~\ref{sec:analysis}, to compare the models with HSPICE results. 
The baseline analytical model considers only linear non-idealities. 
We observe, in Figure \ref{fig:model_v_ana} that even at a low supply voltage of $V_{supply} = 0.25V$, GENIEx achieves a Root Mean Square Error (RMSE) of $0.25$ while estimating the NF with respect to HSPICE. 
This is $7\times$ lower than the baseline analytical model. 
For higher supply voltage of $V_{supply} = 0.5V$, GENIEx achieves a RMSE of $0.7$, which is $12.7\times$ lower than the analytical model. 
To evaluate the impact of non-idealities on large scale DNNs, we develop a functional simulator that incorporates GENIEx with detailed MVM architecture model. 

\section{Functional Simulator} \label{sec:func-sim}

\begin{figure*}[t]
	\centering
	\includegraphics[width=\textwidth, keepaspectratio]{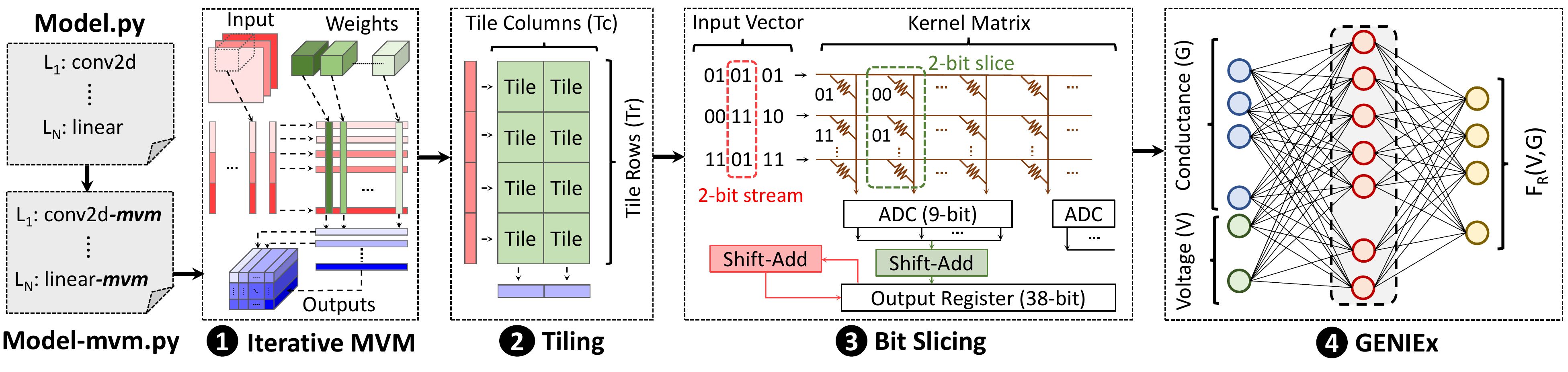}
	\vspace{-6mm}
	\caption{Logical organization of functional simulator}
	\label{fig:func-sim}
	\vspace{-5mm}
\end{figure*}

Several frameworks such as Ares~\cite{reagen2018ares}, Distiller~\cite{neta_zmora_2018_1297430} etc. have been developed using TensorFlow and PyTorch to enable hardware-software codesign studies.
However, such frameworks cannot emulate the implications of crossbar-based hardware, because of the intrinsic differences in the CMOS-based and Crossbar-based computation models.
For CMOS, matrix operations (ops) in a ML model are expressed as General Matrix-Matrix Multiplications (GEMMs) that use floating/fixed point compute units, whereas Crossbar requires matrix ops expressed as Matrix-Vector Multiplications (MVM) that use bit-serial compute units~\cite{shafiee2016isaac, ankit2019puma}.
To address this, we design a functional simulator using PyTorch that implements the \textit{conv2d} (convolution) and \textit{linear} (fully connected) layers based on the crossbar-based computation (\textit{conv2d-mvm, linear-mvm}).

% 2. Crossbar-based compute model steps - each step's behaviour: input, output.
\textbf{Functional Simulator. }
As shown in Figure~\ref{fig:func-sim}, the execution of a convolution layer is divided into three phases within the functional simulator: \textit{Iterative-mvm}, \textit{Tiling}, and \textit{Bit-slicing}.
Each phase depends on parameters that either capture the layer or architecture details pertinent to MVMs.
Consequently, we extract the analog computing aspect of crossbar hardware and ignore any impact of memory and communication.
First, Iterative-mvm expresses a convolution as repeated MVMs, where the weights forms the matrix and a block of pixels across all input channels form the vector (for an iteration).
Each iteration produces an output vector which is comprised of one pixel from all output channels.
Second, Tiling expresses the weight-matrix as a combination of several sub-matrices (or tiles) where, each sub-matrix's size equals the crossbar size.
A slice of input vector is shared by tiles in a row.
Tiles in a column produce partial sums, which are added together to produce a slice of the convolution output.
%Hence, Tiling enables simulating layers of any dimension using practical crossbar sizes.
Third, Bit-slicing (both input and weight bits) expresses the bit-serial nature of crossbar computations~\cite{shafiee2016isaac, ankit2019puma}.
We will refer to a bit-slice ($\ge1$ bits) of inputs and weights as stream and slice, respectively.
Within each step, an input stream is applied to a crossbar's rows to produce ADC outputs.
Next, the shift-and-add units merge the ADC outputs of different weight slices.
Eventually, the outputs of successive input streams go through shift-and-add units to produce the partial sums for a tile.
Depending on the simulation mode (ideal or non-ideal), the ADC outputs are generated either by actual dot-product computation or a forward pass of GENIEx discussed in Section~\ref{sec:nn-model}.
\textit{In summary, the three phases together provide the projection of a layer's execution on actual crossbar hardware}.

% 3. Implementation details - key insights
\textbf{PyTorch Modelling. } The weight-matrix and input-vectors are modelled as multi-dimensional tensors of shape - $(Slices, Tr, Tc,$ $Xr, Xc)$, and $(Batch \ Size, Tr, Xr, Streams)$, respectively. 
Here, the symbols - $T$, $X$, $r$, and $c$ refer to a tile, crossbar, row and column.
Accordingly, $Tr$ refers to a ``tile row" and so on. 
The tensor operations \textit{torch.mul} and \textit{torch.sum} execute the individual crossbar operations.
Subsequently, reduction across the weight slices ($Slices$) and input streams ($Streams$) with scalar factors for shift-and-add generates the partial products.
Subsequently, reduction across $Tr$ dimension produces the convolution output.
Multiple input vectors corresponding to different iterations of MVM are implemented as a batch of vectors ($Batch \ Size$).
Table~\ref{tab:fs-parameters} lists the layer and architecture parameters supported by the functional simulator.

\begin{table}[t]
    %\shrinkAroundTable
    \caption{Functional simulator parameters}
    \shrinkAroundCaption
    \centering
        \resizebox{1\columnwidth}{!}{
            
\begin{tabular}{|l|l|}
\hline
\textbf{Component} & \textbf{Parameters (architecture parameters in \textit{italics})}                                                                                                          \\ 
\hline
\hline
\textbf{Iterative-mvm}   & \begin{tabular}[c]{@{}l@{}}Input feature size, Kernel size, Input channels, \\ Output channels, Padding, Stride\end{tabular} \\ 
\hline
\textbf{Tiling}          & \textit{Crossbar size}                                                                                                                \\ 
\hline
\textbf{Bit-slicing}     & \begin{tabular}[c]{@{}l@{}}Input bits, Weight bits, \textit{Accumulator width} \\ \textit{ADC bits}, \textit{Stream width}, \textit{Slice Width} \end{tabular} \\
\hline
\textbf{GENIEx}        & \textit{Crossbar size}, \textit{R\textsubscript{on}}, \textit{R\textsubscript{off}}, \textit{R\textsubscript{source}}, \textit{R\textsubscript{sink}}, \textit{R\textsubscript{wire}}                                                                                             \\ 
\hline

\end{tabular}

        }
    %\shrinkAroundTable
    \label{tab:fs-parameters}
    \vspace{-7mm}
\end{table}

\section{Experimental Methodology}
\textbf{Crossbar:} We simulate memristive crossbars using HSPICE. The test vectors for $V$ and $G$ are collected from the dataset (CIFAR-100 and ImageNet) and the pretrained neural network models (ResNet) respectively. $I_{non-ideal}$ obtained from SPICE simulations is used to calculate the non-ideality ratio, $f_R$, described in Section \ref{sec:nn-model}. Finally, $V$, $G$ and $f_R(V,G)$ are normalized to the range [0,1] to form the training set for GENIEx. To verify the generalization and applicability of GENIEx, we generated datasets for crossbar configurations with different design parameters such as crossbar size (16, 32, 64), ON resistance (50k$\Omega$, 100k$\Omega$, 300k$\Omega$), and conductance ON/OFF ratio (2, 6, 10). The non-ideality parameters are $R_{source} = 500 \Omega/1000 \Omega$, $R_{sink} = 100 \Omega/500 \Omega$, $R_{wire} = 2.5 \Omega$ per cell. The device parameters are $d_0=0.25nm$, $V_0=0.25V$, $I_0 = 0.1 mA$ \cite{xu2014modeling, yu2012neuromorphic}.

% \begin{table}[t]
%     %\shrinkAroundTable
%     \caption{Design and Non-ideality parameters}
%     \shrinkAroundCaption
%     \centering
%         \resizebox{1.0\columnwidth}{!}{
%             \input{fig/06-methodology/design-param.tex}
%         }
%     %\shrinkAroundTable
%     \label{param}
%     \vspace{-8mm}
% \end{table}
\textbf{Functional Simulator: }The precisions of different components of the functional simulators are as follows: accumulator = 32-bit (24 fractional), ADC = 14-bit, inputs and weights = 16-bit (13 fractional), input Streams = 4-bit, weight Slices = 4-bit, unless otherwise specified. All networks use fixed-point (FxP) representations.

\textbf{DNN: } GENIEx has 500 hidden layer neurons and ReLU non-linearity \cite{nair2010rectified}. We use PyTorch to evaluate large-scale neural networks on the functional simulator using GENIEx.  For the CIFAR-100 dataset, we use the network architecture ResNet-20. For the ImageNet dataset, we considered ResNet-18 on a subset of 7680 test images of the dataset. We report the top-1 accuracies for both datasets. The ideal floating point 32-bit (FP) accuracies for CIFAR-100 and subset of ImageNet are 69.6\% and 76.01\% respectively.

%\shrinkAroundFigure

\section{Results} \label{sec:results}

\subsection{Impact on Design Parameters}

First, we study the impact of non-idealities on the classification accuracy of DNNs under different design considerations of crossbar sizes, ON resistances, and conductance ON/OFF ratio. 
The studies are performed on ResNet-20 for CIFAR-100 dataset with the features of bit-slicing and bit-streaming using 4-bit Streams and Slices. 
The weights and activations for these networks have been considered as 16-bit fixed point representations.

\begin{figure}[t]
	\centering
	\includegraphics[width=\columnwidth, keepaspectratio]{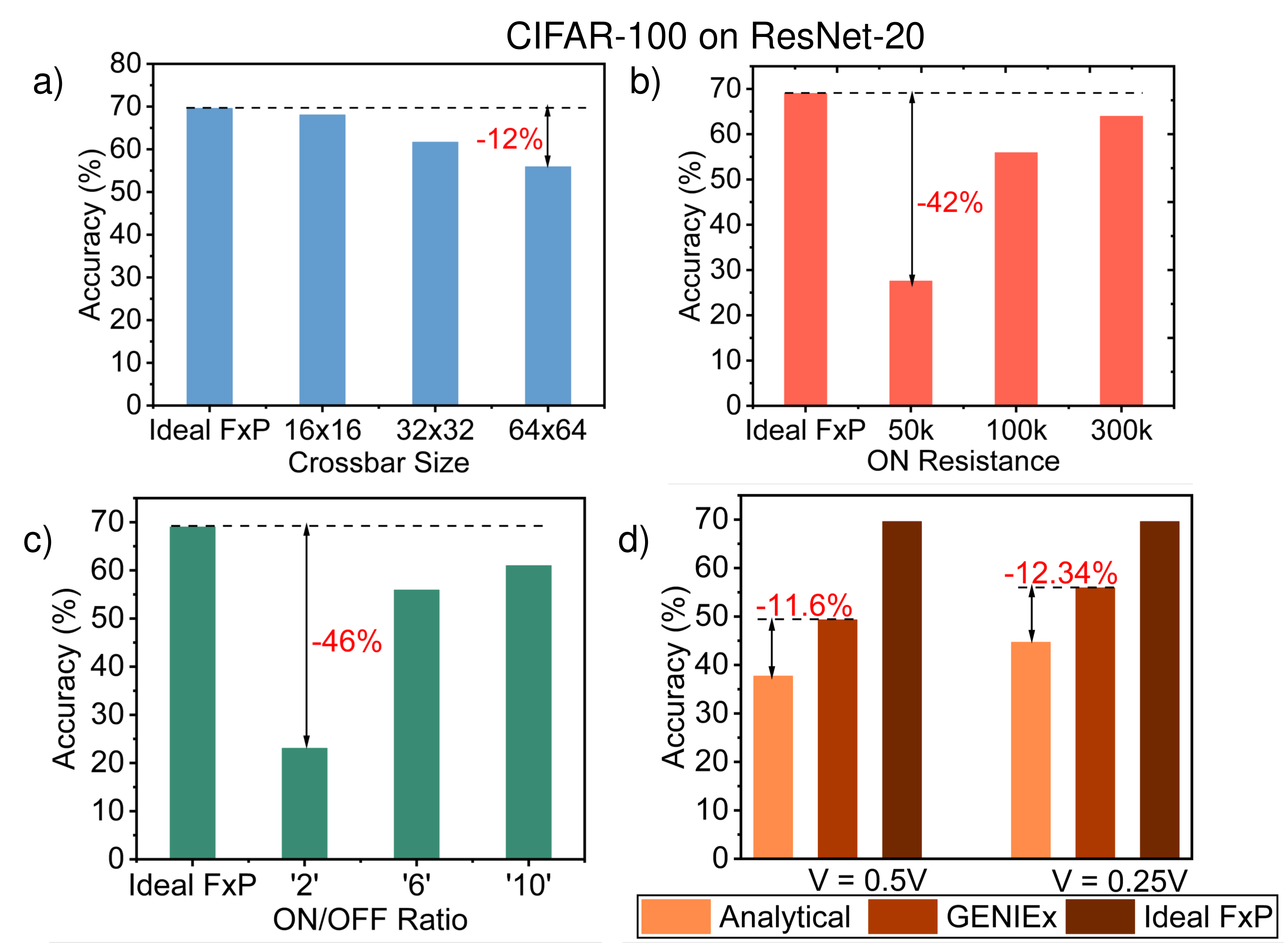}
    \shrinkAroundFigure
	\caption{Impact of non-idealities with crossbar design parameters (a) Crossbar Size, (b) ON resistance, (c) ON/OFF ratio. (d) Comparison between analytical model and GENIEx. }
	\label{fig:results_fig1}
	\shrinkAroundFigure
\end{figure}

We observe in Figure~\ref{fig:results_fig1} (a) that the classification accuracy degrades by $12\%$ for a $64\times64$ crossbar compared to an ideal 16 bit fixed-point (Ideal FxP) implementation.
However, for lower crossbar sizes like $16\times16$, the degradation is $\le1\%$.
This is due to reduced effective resistance for higher crossbar sizes, as discussed in Section~\ref{sec:analysis}.
For higher ON resistances such as 300k$\Omega$, we observe, in Figure~\ref{fig:results_fig1} (b) that the accuracy degradation is $7.6\%$ lower than the case with $100k\Omega$.
This is because the parasitic resistances have more pronounced effect on crossbars with lower ON resistances, resulting in higher accuracy degradation.
In~\ref{fig:results_fig1} (c), we observe that for a given ON resistance ($100k\Omega$ in this case), lower ON/OFF ratio like $2$ results in upto $46\%$ degradation in accuracy due to the average resistances in the crossbar being low.
With higher ON/OFF ratio such as $10$, the accuracy degradation reduces to $8.6\%$. 

Figure~\ref{fig:results_fig1} (d) shows that there is a significant difference between the accuracies predicted by an analytical model and GENIEx.
Further, it also illustrates that an analytical model overestimates the accuracy degradation by $12.34\%$ for supply voltage, $V_{supply} = 0.25V$ and $11.6\%$ for $V_{supply} = 0.5V$ compared to GENIEx. 
Note that the analytical model considers only linear non-idealities (parasitic resistances).
\textit{It underscores that the device non-linearity which is captured by our model can push the behavior of the crossbar towards ideality, thus resulting in a lower accuracy degradation.}

\begin{figure}[t]
	\centering
	\includegraphics[width=\columnwidth, keepaspectratio]{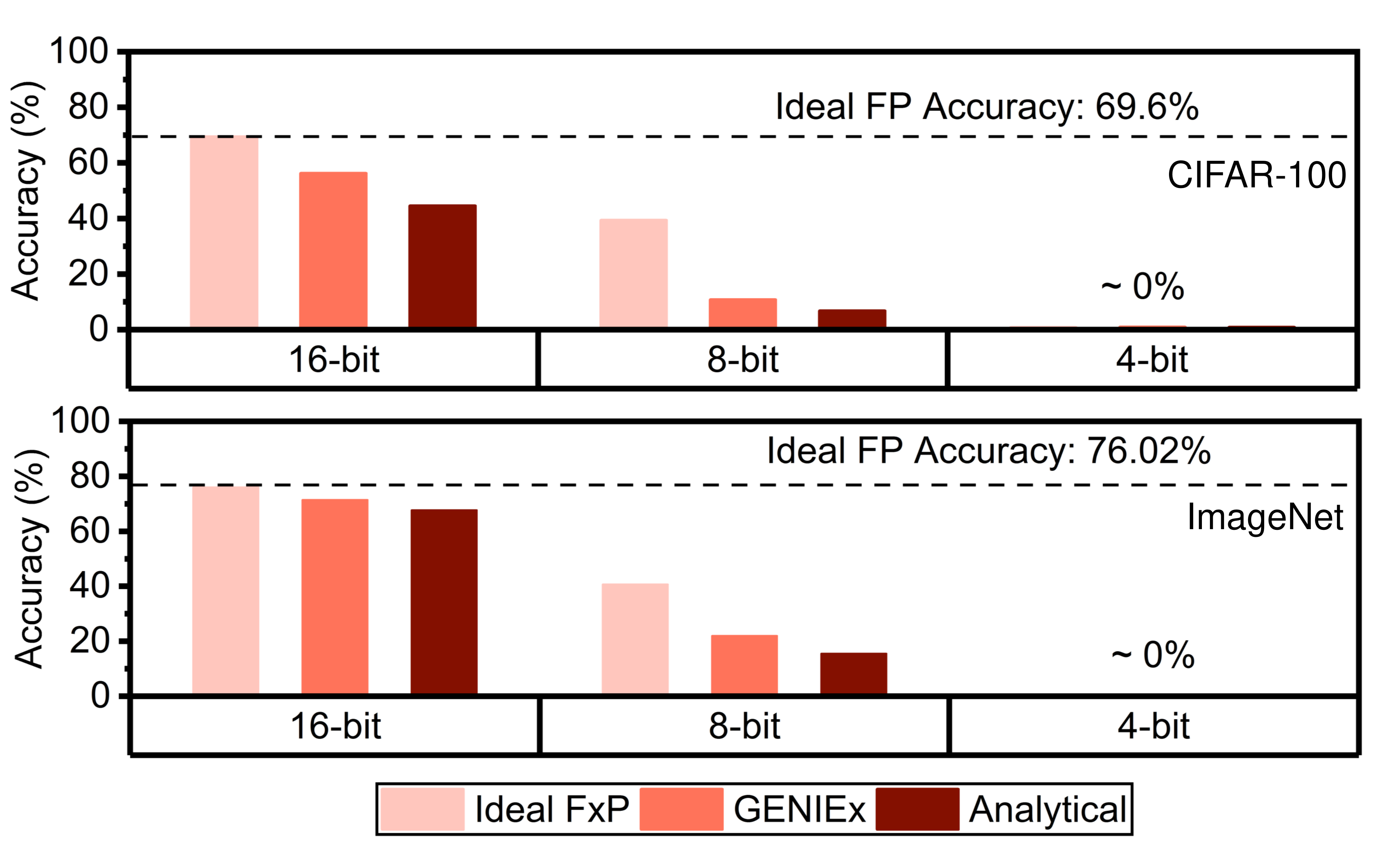}
\shrinkAroundFigure
	\caption{Impact of precision of weights and activations on classification accuracy under the influence of non-idealities.}
	\label{fig:results_fig2}
	\shrinkAroundFigure
\end{figure}

\subsection{Impact of Quantization} \label{sec:results-quantization}

We study the effect of non-idealities on DNNs with different bit-precision for weights and activations. 
We consider 3 cases for networks where the weights and activations are 16-bit, 8-bit and 4-bit fixed point representations: i) Ideal, ii) Non-idealities estimated by analytical model, and iii) Non-idealities estimated by GENIEx. 
Figure~\ref{fig:results_fig2} shows that when the weights and activations are represented as 16-bit fixed point, the classification accuracy is close to ideal 32-bit floating point accuracy.
When the precision of the weights and activations is reduced to 8-bit, the accuracy degrades by $30.03\%$ for CIFAR-100, and $35.29\%$ for ImageNet.
For 4-bit case, the accuracy is~$\simeq0\%$. 
Further, the accuracy degradation increases from $12.5\%$ to $29.6\%$ for CIFAR-100 and $4.54\%$ to $17.67\%$ for ImageNet when the bit-precision is reduced from $16$-bit to $8$-bit.
Thus, non-idealities have an increased detrimental effect at lower bit-precisions.
Note that the analytical models overestimate the degradation in accuracy by $12.34\%$ and $3.99\%$ for CIFAR-100, and $3.70\%$ and $6.49\%$ for ImageNet compared to GENIEx for 16-bit and 8-bit cases respectively. 
\textit{This result shows that due to the constraints on a) the number of bits NVM devices can store reliably~\cite{Hu2018MNIST}, and b) the DAC precisions for efficient MVM~\cite{shafiee2016isaac}, bit slicing of weights and inputs are essential features to achieve close to full-precision accuracies.}

\begin{figure}[t]
	\centering
	\includegraphics[width=\linewidth, keepaspectratio]{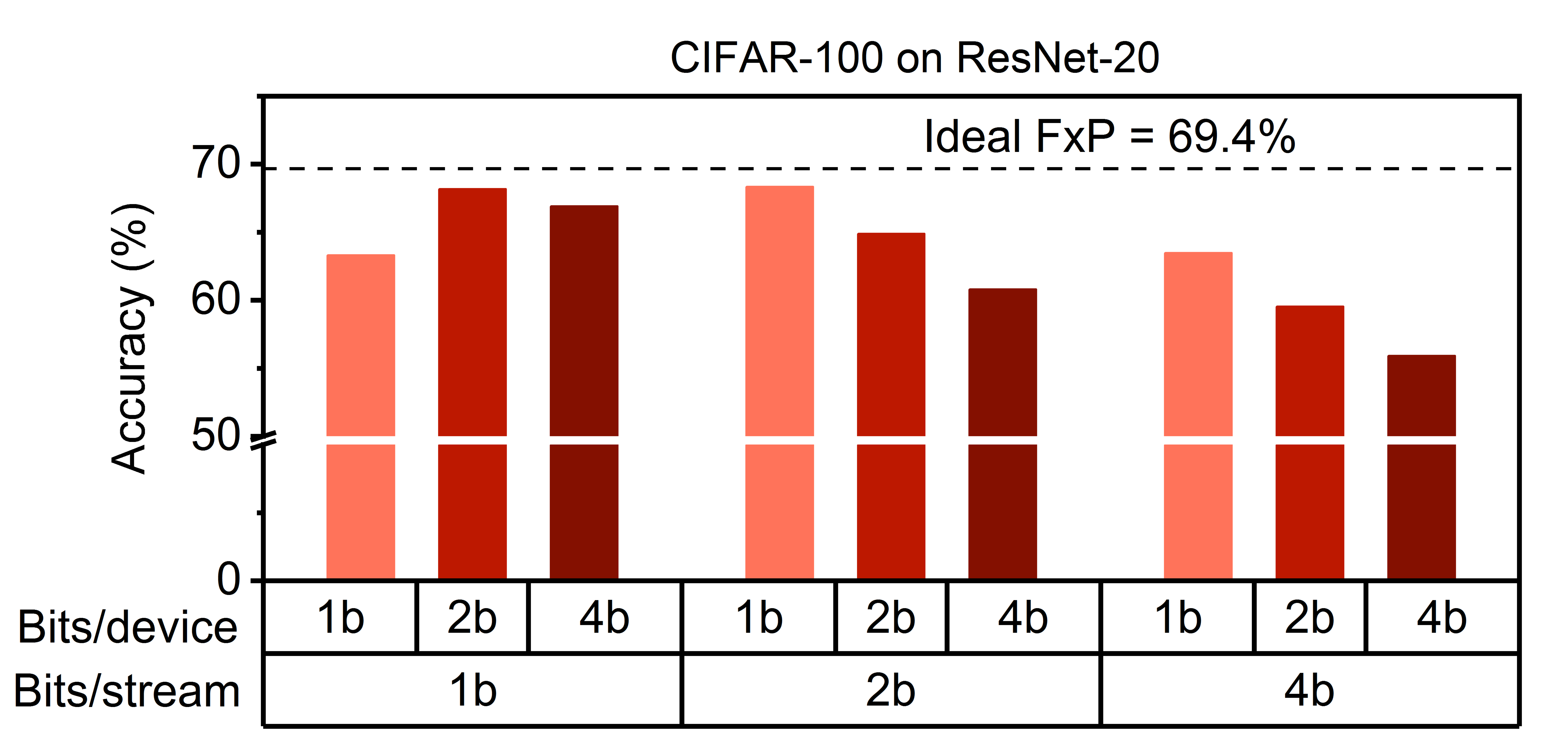}
    \shrinkAroundFigure
	\caption{Impact of number of bits/device and bits/stream.}
	\label{fig:results_fig3}
	\shrinkAroundFigure
\end{figure}

\subsection{Impact of Bit Slicing}

Finally, we study the impact of different bit-slicing configurations for inputs (Streams) and weights (Slices) for 16-bit FxP network on the classification accuracy of DNNs in presence of non-idealities. 
Figure~\ref{fig:results_fig3} shows that using 2-bit or 1-bit Streams and Slices, achieves close to ideal FxP accuracy.
Increasing the Stream and Slice widths to 4-bit results in a $12.48\%$ degradation in accuracy.
Note that 1-bit Streams and 1-bit Slices result in a slightly lower accuracy. 
This is because the combination of 1-bit Streams and Slices results in very high sparsity that makes the crossbar resilient to parasitic resistances. 
In such a case, device non-linearity can lead to non-ideality factor, $NF$ to be lower than $0$, resulting in lower accuracy. 
Nonetheless, using lower number of bits per slice and stream can help achieve close to ideal FxP accuracies. 
\textit{This result provides a perspective on architectural design parameters such as the Slice and Stream widths in presence of crossbar non-idealities.}

\section{Conclusion}

We present GENIEx, a generalized approach to emulating non-ideality in memristive crossbars using neural networks.
We perform extensive SPICE simulations and subsequently train a neural network to learn a generalized behavior of the non-ideal crossbar. Finally, we use GENIEx in a functional simulator for evaluating the impact of these non-idealities on the image classification performance of large-scale DNNs. 
We show that GENIEx achieves a \textit{low} RMSE of $0.25$ for $V_{supply} = 0.25V$ and $0.7$ for $V_{supply} = 0.5V$ with respect to HSPICE, which is $7\times$ and $12.8\times$ lower, respectively, than an analytical model. 
This is due to the ability of GENIEx to model both linear and non-linear non-idealities. 
We further show that an analytical model overestimates the degradation in classification accuracy by $12.3\%$ on CIFAR-100, and $4\%$ on ImageNet compared to GENIEx. 
We analyze the impact of non-idealities on the crossbar design parameters such as crossbar-size, ON resistance, conductance ON/OFF ratio, Stream width, and Slice width. 
We observe that packing lower bits per device as well as using low crossbar sizes with higher ON resistances is necessary to minimize the impact of non-idealities. The proposed end to end framework for evaluating crossbar based architectures on realistic crossbars can pave the way for efficient crossbar designs for future machine learning systems.

\vspace{-2mm}
%\clearpage
\section*{Acknowledgements}
This work was supported in part by the Center for Brain-inspired Computing Enabling Autonomous Intelligence (C-BRIC), one of six centers in JUMP, a Semiconductor Research Corporation (SRC) program sponsored by DARPA, in part by the National Science Foundation, in part by Intel, and in part by the Vannevar Bush Faculty Fellowship.

\balance
\bibliographystyle{unsrt}

\end{document}